\documentclass{article}

\pdfoutput=1
\usepackage{PRIMEarxiv}

\usepackage[utf8]{inputenc} 
\usepackage[T1]{fontenc}    
\usepackage{hyperref}       
\usepackage{url}            
\usepackage{booktabs}       
\usepackage{amsfonts}       
\usepackage{nicefrac}       
\usepackage{microtype}      
\usepackage{lipsum}
\usepackage{fancyhdr}       
\usepackage{graphicx}       
\graphicspath{{media/}}     

\usepackage{changepage}

\usepackage{afterpage}

\usepackage{textcomp,marvosym}
\usepackage{scalerel}

\usepackage{amsmath,amssymb}
\usepackage[normalem]{ulem}

\usepackage{algorithm}
\usepackage{algpseudocode}

\usepackage{array}

\newcolumntype{+}{!{\vrule width 2pt}}

\newlength\savedwidth

\newcommand\thickhline{\noalign{\global\savedwidth\arrayrulewidth\global\arrayrulewidth 2pt}%
\hline
\noalign{\global\arrayrulewidth\savedwidth}}

\title{Optimal non-pharmaceutical intervention policy for Covid-19 epidemic via neuroevolution algorithm

}

\author{
  Arash Saeidpour \\
  Center for the Ecology of Infectious Diseases \\
  Odum School of Ecology, University of Georgia\\
  Athens, GA 30602\\
  \texttt{arashs@uga.edu} \\
   \And
  Pejman Rohani \\
  Center for the Ecology of Infectious Diseases \\
  Department of Infectious Diseases \\
  Odum School of Ecology, University of Georgia\\
  Athens, GA 30602\\
  \texttt{rohani@uga.edu} \\
}

\begin{document}
\maketitle

\begin{abstract}
National responses to the Covid-19 pandemic varied markedly across countries, from business-as-usual to complete shutdowns. Policies aimed at disrupting the viral transmission cycle and preventing the healthcare system from being overwhelmed, simultaneously exact an economic toll. We developed a intervention policy model that comprised the relative human, economic and healthcare costs of non-pharmaceutical epidemic intervention and arrived at the optimal strategy using the neuroevolution algorithm.  The proposed model finds the minimum required reduction in contact rates to maintain the burden on the healthcare system below the maximum capacity. We find that such a policy renders a sharp increase in the control strength at the early stages of the epidemic, followed by a steady increase in the subsequent ten weeks as the epidemic approaches its peak, and finally control strength is gradually decreased as the population moves towards herd immunity. We have also shown how such a model can provide an efficient adaptive intervention policy at different stages of the epidemic without having access to the entire history of its progression in the population. This work emphasizes the importance of imposing intervention measures early and provides insights into adaptive intervention policies to minimize the economic impacts of the epidemic without putting an extra burden on the healthcare system.
\end{abstract}


\section{Introduction}
On March 11, 2020 the World Health Organization (WHO) announced that Covid-19, caused by severe acute respiratory syndrome coronavirus 2 (SARS-CoV-2) \cite{zhu2020novel}, "can be characterized as a pandemic" \cite{WHOpandemic}. Within a month, most countries around the world had taken public health measures to contain the spread of the novel virus \cite{govresponse}. However, the type and severity of implemented measures and their subsequent success in minimizing the public health impacts of the outbreak varied greatly by country \cite{brauner2021inferring}. This variation in policies and their effectiveness reflects the complexity of finding the balance between two often competing policy objectives:  protecting the public's health versus minimizing the economic impact of intervention measures\cite{brett2020transmission}.\\

Initially, without access to pharmaceuticals, studies focused on two distinct control approaches:  mitigation and suppression \cite{prem2020effect,walker2020impact,davies2020effects}. The mitigation strategy aims to reduce transmission such that healthcare systems are not overwhelmed, while aiming to maintain the chain of transmission in order to achieve herd immunity. In contrast, the suppression strategy is aimed at virus elimination.  In hindsight, countries that acted early to suppress the disease have excelled at minimizing both the public health and economic impact of the epidemic \cite{hassan2021hindsight,dong2020interactive,kochanczyk2021pareto}. While early suppression measures appear to outperform the mitigation strategy both in terms of public health goals and economic costs, such policies would not necessarily be successful in countries where citizens are more averse to government-enforced control and surveillance measures \cite{anderson2020will}. Moreover, suppression measures would only be successful if implemented in the early stages of the epidemic and sufficiently strictly as to curtail transmission effectively.  In a number of settings, however, suppression has been implemented in a piece-meal manner, leading to periods of drastic interventions including lockdowns punctuated by relaxation of social distancing measures and subsequent uptick in  transmission \cite{10.1038/s41467-021-22366-y,Hollingsworth:2011jr}.  This prompted us to examine the optimal mitigation strategy, which aims to manage or mitigate the healthcare impacts of the epidemic while population approaches herd immunity.\\

Characterizing immediate and long-term economic, social and human burden of Covid-19 epidemic is  challenging and has led to several research efforts to examine the optimal intervention policy from various perspectives. It is unfeasible to review comprehensively this body of work, so we confine ourselves to a number of the key studies. Rowthorn and Maciejowski~\cite{rowthorn2020cost} investigated the optimal uniform lockdown in an $SIR$ model assuming a variety of parameterizations \cite{rowthorn2020cost}. Their objective function assigned monetary values to costs arising from infection, lockdown, and value of life. Their main finding  was that in the medium term, a policy that maintains effective reproduction number value close to 1 provides the best path. Bethune and Korinek~\cite{bethune2020covid} contrasted the decisions made by rational, individual agents with the choices made by a social planner who is able to coordinate the choices of individuals \cite{bethune2020covid}. They found that rational agents generate large externalities because they fail to internalize the effects of their economic and social activities on others' risk of infection. Alvarez \textit{et al.} formalized the social planner's dynamic control using an $SIR$ epidemiological model and a linear economy. The best strategy starts with a severe lockdown two weeks after the epidemic, covers 60\% of the population after a month, and progressively decreases to 20\% of the population after three months.  More recently, a number of  studies have broadened this  exploration to identify age-specific optimal control strategies \cite{acemoglu2020optimal,richard2021age}.\\

Inspired by \cite{SalimansEvolutionLearning,SuchDeepLearning,Riolo&Rohani:2015,davoodi2021graph,faryadi2021reinforcement}, we sought to use an neuroevolution strategy to finding the optimal policy function which would dynamically determine the minimal required reduction in transmission rates at each time instant, deemed as \textit{control strength} hereafter. Reductions in transmission may result from lower contacts (due to isolation-in-place ordinances, movement restrictions, or lockdown policies), or the adoption of personal protective measures that serve to curtail transmission upon contact (such as the use of face masks), with varying economic impact. The fitness function is expressed such that a strategy is rewarded for allowing the epidemic to  remove individuals from the susceptible pool without overwhelming the healthcare capacity. The proposed neuroevolution strategy begins by initializing a population of random policy functions. The generated policy functions are then used to simulate the trajectory of the epidemic. The fitness of each function is then evaluated based on a reward function. The most elite policy functions are then perturbed (mutated) to generate the next generation offsprings. The new population is then evaluated and this process is repeated for a pre-defined number of iterations. We also derived the optimal control solution via Pontryagin's maximum principle (PMP) \cite{pontryagin1987mathematical} and compared the results with neuroevolution optimal policy. \\ 

\begin{figure}[!h]
\centering
\includegraphics[width=\linewidth]{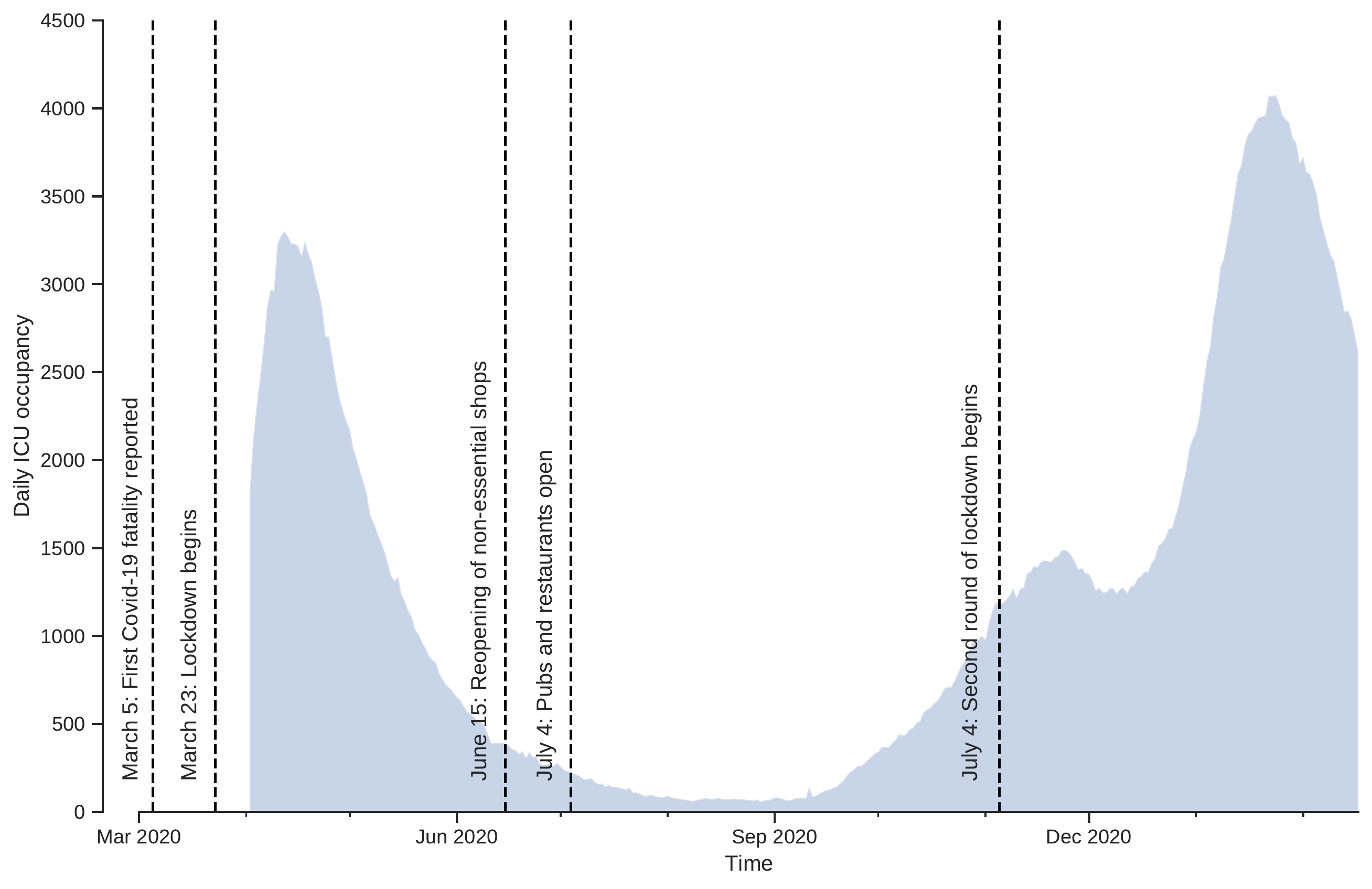}
\caption{{\bf Number of Covid-19 patients in intensive care (ICU) and timeline of lockdowns in the UK.}
}
\label{timeline}
\end{figure}

We have chosen the United Kingdom as our target population to implement the proposed approach. The choice of the UK as our target population was largely motivated by the frequent changes in the government's strategy to contain the epidemic \cite{cameron2020variation}, as summarized in Figure \ref{timeline}. The UK's initial response was a mitigation policy, majorly inspired by the response to the flu pandemic, with an emphasis on protecting the most vulnerable to avoid overburdening the healthcare system in an effort to achieve herd immunity \cite{hassan2021hindsight}. This initial policy later changed to a suppression policy by implementing lock-downs and imposing face mask-wearing requirements. Looking back at the early days of the epidemic, this study aims to understand how an effective mitigation policy could have been implemented (see \cite{hassan2021hindsight} for a comparison of initial responses to Covid-19 by different countries including United Kingdom).\\

Our study explores mechanisms for "flattening the curve" -- it is motivated by COVID-19 pandemic but need not be restricted to precise courses of action undertaken in the response to pandemic.  Our findings are intended to be informative for future epidemic control, particularly at early stages of epidemic where no effective vaccine is in sight. 

The ideal intervention policy results in a rapid increase in control strength early in the epidemic, followed by a sustained increase over the next ten weeks as the epidemic reaches its peak, and ultimately a progressive drop in control strength as the population achieves herd immunity. We've also shown how, without having access to the complete history of the epidemic's growth in the population, such a model may give an effective adaptive intervention policy at various stages of the epidemic. This study highlights the significance of implementing control measures as promptly as possible and offers insights into adaptive intervention strategies aimed at reducing the economic effect of epidemics while avoiding undue strain on the healthcare system.

\section{Materials and methods}

\subsection{Model structure}
We used a deterministic, time-varying Susceptible-Exposed-Infectious-Recovered-Hospitalized in ICU ($SEIRH$) model \cite{keeling&Rohani:2008,borchering2021anomalous} to characterize the transmission dynamics in the UK as described in Eqs.~\ref{eqn:1}--\ref{eqn:6}: 

\begin{eqnarray}
\label{eq:SEIR_1}
\dot{S} = \frac{\mathrm{d}S}{\mathrm{d}t} &=& - (1 - c(t)) \frac{\beta SI}{N} \label{eqn:1} \\
\dot{E} = \frac{\mathrm{d}E}{\mathrm{d}t} &=& (1 - c(t)) \frac{\beta SI}{N} - \rho E  \\
\dot{I} = \frac{\mathrm{d}I}{\mathrm{d}t} &=& \rho E - \gamma I -P_{\scaleto{Detection}{3pt}} \sigma_{\scaleto{ICU}{3pt}} \gamma_{\scaleto{ICU Delay}{3pt}} I \\
\dot{R} = \frac{\mathrm{d}R}{\mathrm{d}t} &=& \gamma  I + \gamma_{\scaleto{ICU Stay}{3pt}} H \\
\label{eq:SEIR_5}
\dot{H} = \frac{\mathrm{d}H}{\mathrm{d}t} &=& P_{\scaleto{Detection}{3pt}} \sigma_{\scaleto{ICU}{3pt}} \gamma_{\scaleto{ICU Delay}{3pt}} I -  \gamma_{\scaleto{ICU Stay}{3pt}} H  \label{eqn:6}
\end{eqnarray}

where $\beta$ is the transmission rate, $1/\rho$ and $1/\gamma$ give the mean latent and infectious periods, respectively and   $c(t) \in [0,1]$ is the reduction in transmission (such that $c(t)=1$ signifies complete cessation of transmission).  The state variable $H(t)$ denotes the number of occupied ICU beds and is determined by the probability that an infection is detected ($P_{\scaleto{Detection}{3pt}}$), the fraction of cases that require ICU treatment ($\sigma_{\scaleto{ICU}{3pt}}$) and the rate of admission to the ICU ($\gamma_{\scaleto{ICU Delay}{3pt}}$). The mean duration of stay in the ICU is determined by $1/\gamma_{\scaleto{ICU Stay}{3pt}}$. Model parameters and chosen values  are presented in Table \ref{table1}.   \\

In our analyses, we examine changes in optimal intervention policy assuming policies are implemented starting at different points during the epidemic, $T_0$.  To identify the appropriate initial conditions at these different starting points, we used a particle filter \cite{Shrestha:2011ho} to estimate the effective retrospective daily $c(t)$ (where $t=0,\ldots,T_0$), thus yield the epidemiological state of the population at different stages of the epidemic. The agreement between our fitted $SEIRH$ model and data is shown in Figure~S2.  

\begin{table}[!ht]
\begin{adjustwidth}{-0.0in}{0in} 
\centering
\caption{
{\bf Parameters of SEIRH model}}
\begin{tabular}{l p{0.45\linewidth} l l}
{\bf Parameter} & {\bf Definition} & {\bf Value} & {\bf Source} \\ \thickhline
\(N\) & Total population size & 66,436,000 & \cite{park2020population}\\
\(R_0\) & Basic reproduction number & 2.3 & \cite{li2020early,zhang2020evolving} \\
\(1 / \gamma\) & Mean infectious period (days) & 2.9 & \cite{li2020early,zhang2020evolving} \\
\(1 / \rho\) & Mean latent period (days) & 3.4 & \cite{li2020substantial} \\
\(\beta\) & Mean transmission rate (1/day) & 0.793 & Estimated \\
\(P_{\scaleto{Detection}{3pt}}\) & Ratio of confirmed cases to total infections & 0.3 & \cite{giattino2020epidemiological} \\
\(\sigma_{\scaleto{ICU}{3pt}}\) & Proportion of confirmed cases that end up in ICU & 0.05 & \cite{UK_ICU} \\
\(1 / \gamma_{\scaleto{ICU Delay}{3pt}}\) & Median time from symptoms onset to ICU admission (days) & 10 & \cite{wang2020clinical} \\
\(1 / \gamma_{\scaleto{ICU Stay}{3pt}}\)  & Mean ICU stay period (days) & 9 & \cite{grasselli2020baseline} \\

\(H_{max}\) & Number of ICU beds & 4074 & \cite{UKICUbeds} \\
\hline
\end{tabular}
\begin{flushleft} The table presents the parameters of SEIRH model used to model the dynamics of Covid-19 transmission in the population of UK.
\end{flushleft}
\label{table1}
\end{adjustwidth}
\end{table}

\section{Reward function}
We first introduce the following multi-objective reward function to account for three  opposing goals: i) Sustain viral transmission  to achieve herd immunity, ii) Keep the ICU occupancy below the maximum capacity, and iii) Impose the minimum possible control:

\begin{equation} \label{r1}
\begin{split}
r_1(t) &= \alpha_1 r_1(t)_{Herd \; Immunity} - \alpha_2 r_1(t)_{Exceedance} -\alpha_3 c(t)^2 \\ 
& = \alpha_1 E(t)/N - \alpha2 (H(t)-H_{max})/H_{max}- \alpha_3 * c(t)^2. \\
\end{split}
\end{equation}

We defined $r_1(t)$ for the sake of mathematical simplicity in deriving PMP solution and it is only used to compare the optimal NPI policies obtained from neuroevolution and PMP methods. For the remainder of this study, we use a slightly different objective function, $r_2(t)$, defined as follows:

\begin{equation} \label{r2}
\begin{split}
r(t) & = \alpha_1 r_{{\scaleto{\mbox{Herd Immunity}}{5pt}}}(t) + \alpha_2 r_{{\scaleto{\mbox{Exc}}{4pt}}}(t) + \alpha_3 r_{{\scaleto{\mbox{Control}}{4pt}}}(t), \\
 & = \alpha_1 (R(t)/N) - \alpha_2 Relu((H(t) - H_{max})/N) - \alpha_3 * c(t).
\end{split}
\end{equation}

In both reward functions (equations (6) \& (7)), the  terms $\alpha_1$, $\alpha_2$ and $\alpha_3$ modulate the relative importance of herd immunity, healthcare burden and economic costs, respectively. The goal, therefore, is to identify the optimal intervention function $c(t)$ that maximizes the sum of rewards, $J$, during the course of epidemic:\\

\begin{equation} \label{sum_rew}
\max_{c(t)} J = \int r_i(t) dt, i \in {1,2} 
\end{equation}

\section{Pontryagin's maximum principle (PMP)}
In this section we first derive the necessary conditions for optimal control via Pontryagin's maximum principle, and describe the iterative numerical algorithm (the forward-backward sweep method) used to find the optimal solution. First, we form the following Hamiltonian function:

\begin{equation}
\mathcal{H}(t,\mathfrak{s}(t),c(t),\lambda_\mathfrak{s}(t)) = r(t) + \lambda_S(t) \dot{S} + \lambda_E(t) \dot{E} + \lambda_I(t) \dot{I} + \lambda_R(t) \dot{R} + \lambda_H(t) \dot{H}.
\end{equation}

$\lambda_\mathfrak{s}(t)$ are adjoint functions satisfying the adjoin system:

\begin{eqnarray}
\label{eq:adjoints_1}
\dot\lambda_\mathfrak{s}(t) &=& -\frac{\partial \mathcal{H}(t,\mathfrak{s}^*(t),c^*(t),\lambda^*_\mathfrak{s}(t))}{\partial \mathfrak{s}}, \mathfrak{s} \in \{S,E,I,R,H\}, \\
\label{eq:transvers}
\lambda_\mathfrak{s}(T) &=& 0 \; \text{(Transversality condition)}.
\end{eqnarray}

Expanding equation \ref{eq:adjoints_1} yields:
\begin{align}
\label{eq:adjointsexpanded}
\dot\lambda_S(t) = -\partial{\mathcal{H}} / \partial{S(t)} =& (\lambda_S-\lambda_E)\frac{(1-c) \beta I}{N} \\
\dot\lambda_E(t) = -\partial{\mathcal{H}} / \partial{E(t)} =& (\lambda_E-\lambda_I)\rho - \frac{\alpha_1}{N}\\
\dot\lambda_I(t) = -\partial{\mathcal{H}} / \partial{I(t)} =& (\lambda_E-\lambda_S) \frac{(1-c)\beta SI}{N} + (\lambda_I - \lambda_R) \gamma + \nonumber \\
&(\lambda_I - \lambda_H) \gamma_{\scaleto{ICU Delay}{3pt}} P_{\scaleto{Detection}{3pt}} \sigma_{\scaleto{ICU}{3pt}}\\
\dot\lambda_R(t) = -\partial{\mathcal{H}} / \partial{R(t)} =& 0 \\
\label{eq:adjoints_5}
\dot\lambda_H(t) = -\partial{\mathcal{H}} / \partial{H(t)} =& (\lambda_H - \lambda_R) \gamma_{\scaleto{ICU Stay}{3pt}} + \frac{\alpha_2}{H_{max}}
\end{align}

The necessary conditions for the optimal control is obtained by maximizing the $\mathcal(H)$ with respect to $c(t)$:

\begin{equation}
\frac{\partial\mathcal{H}}{\partial c}=0 \;\; \text{at} \;\; c^*{t} \rightarrow c^*(t) = (\lambda_S-\lambda_E)\frac{\beta I}{2 \alpha_3 N}, c^*(t) \in [0,1]
\end{equation}

The state equations (equations \ref{eq:SEIR_1}-\ref{eq:SEIR_5}) and adjoint equations (equations \ref{eq:adjoints_1}-\ref{eq:adjoints_5}) together with state initial conditions and transversality conditions (equation \ref{eq:transvers}) form the \textit{Optimality system}. The explicit solution can not be analytically derived. Thus we turned to an iterative numerical method, \textit{Forward-backward Sweep}, to solve the \textit{Optimality system}.

\subsection{Neuroevolution algorithm}
\begin{figure}[!h]
\centering

\includegraphics[width=\linewidth]{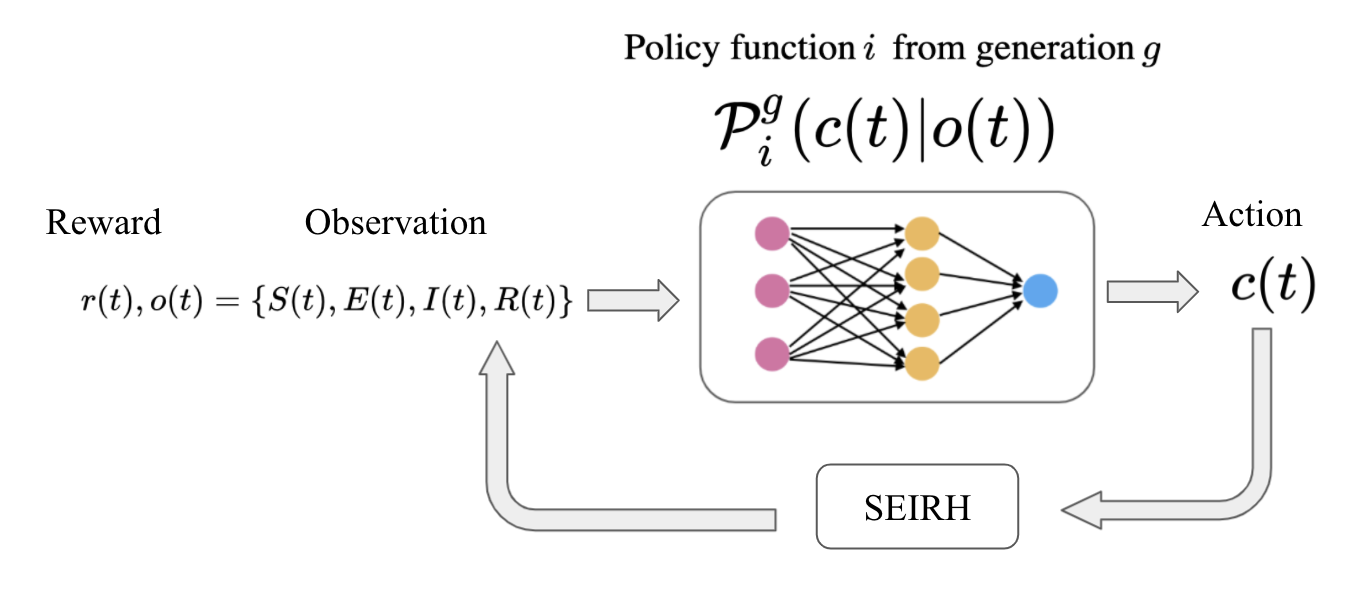}
\caption{{\bf Schematic representation of policy function}
$\mathcal{P}^g_i$, represents the policy function $i$ of generation $g$. The $L$ most elite policy functions of each generation are mutated to generate the $M$ policy functions of next generation.}
\label{schematic}

\end{figure}

The optimal policy function, $\pi_\theta$, is a feed forward neural network, parameterized by $\theta$ which takes the  state of the system at current time $t$, $\{S(t),E(t),I(t),R(t)\}$ as input and returns the control strength, $c(t)$. The neuroevolution strategy aims to find the optimal policy function, $\mathcal{P}_{\text{Most elite}}^{G}$, with highest fitness score. Fitness score of policy function $j$ in generation $i$, $f_j^i$, is equal to the sum of rewards, $J$ (equation \ref{sum_rew}) and is obtained by running the $SEIRH$ model with the corresponding policy function. First, $M$ policy functions ($\mathcal{P}^1_j$) are randomly initialized. For each policy function, a trajectory is rolled out and fitness score is calculated at the end of simulation, as shown in figure \ref{schematic}. The $L$ policy functions with the highest fitness scores are mutated to generate the next generation of policy functions. Mutation is implemented by adding a random Gaussian noise, scaled by the mutation rate, $\sigma$, to $\theta$ parameters of elite policy functions. The new offspring policy functions served as the parents of next generation. This process continues to find a policy function with a sufficiently high fitness score, $\mathcal{P}_{\text{Most elite}}^{G}$. We used a fully-connected feed-forward network with 3 16-unit hidden layers and one tanh output layer to model the policy function. Pseudocode for the neuroevolution algorithm used in this study is provided in Algorithm~\ref{alg:cap}.\\

\begin{algorithm}
\caption{Neuroevolution algorithm}\label{alg:cap}
\begin{algorithmic}
\Require Population size $M$, Number of generations $G$, Elite population size $L$, Mutation rate $\sigma$
\\
\textbf{Initialize} $M$ policy functions, $\mathcal{P}_j^1$, with random initial weights $\theta_j^1$  
\For i=1 to $G$.   \# Iterate G generations
\For j=1 to $M$
\State $f_j \gets$ Roll out a trajectory by running the model using $\mathcal{P}_i^i$ \# Fitness score
\EndFor
\State Sort $\theta^i_j$ by $f_j$ in descending order
\State $\theta_{Elite}^i = \{\theta_j^i | j<L\} \cup \theta_{\text{}{Most \; elite}}^{i-1}$
\For j=1 to $M$
\State Draw sample $t \sim U(1,L)$ \# Select a parent
\State Draw sample $\epsilon \sim \mathcal{N}(0,1)$ \# Gaussian noise
\State $\theta_{j}^{i+1} = \theta_t^i + \sigma \epsilon$ \# Mutate
\EndFor
\EndFor \\
\Return $\mathcal{P}_{\text{Most elite}}^{G}$

\end{algorithmic}
\end{algorithm}

\section{Results}
\subsection{Which optimization algorithm?}
We  compared the optimal intervention policies obtained from PMP and neuroevolution policies (Fig S1). The policies are obtained using the $r_1$ reward function (equation 6) with ${\alpha_2=1e-1,\alpha3=5e-3}$ and same initial conditions. We found the optimal policies obtained from both methods to be very similar. In simpler problems where an analytic solution can be obtained for the optimality system, the PMP method can provide more insights about the optimal control solution and the dynamics of the system.  Otherwise, a neuroevolutionary approach is computationally advantageous since the resulting policy function  provides an optimal strategy for a broad range of initial conditions at a substantially smaller computations cost.  That is, the PMP optimal intervention for a given initial condition is obtained by solving the boundary-value problem formulated in equations (\ref{eq:SEIR_1}-\ref{eq:SEIR_5}) and (\ref{eq:adjoints_1}-\ref{eq:adjoints_5}). For a new boundary condition, the numerical solution must be repeated to solve the new boundary-value problem.  In the remainder of the paper, our optimal solutions are obtained via the neuroevolutionary approach.

\subsection{Reward function exploration}

 The relative economic burden of different objectives in the reward function is determined by the  weights, $\{\alpha_1,\alpha_2,\alpha_3$\}. Thus, we examined the effects of variation in these parameters on the resulting optimal policy ({see Figure~S3}). We constrained $\alpha_1$ to be 1 and changed the values of $\alpha_2$ and $\alpha_3$  over a logarithmic grid. For each parameter set, we trained the neuroevolution algorithm for 2000 generations with a population size of 256.  The resulting policy functions (purple lines) and corresponding ICU occupancy trajectories of the 10 best-performing agents for each parameter set are depicted in Figure S3. We found the reward function to be consistently robust to variation in the values of $\alpha_2$. That is, the tested range of $\alpha_2$ values makes the cost of ICU overflow sufficiently prohibitive, leading to high-fitness strategies ensuring  ICU maximum capacity is not exceeded (note that the ICU overflow reward is equal to 0 while the ICU occupancy is below the maximum capacity and negative otherwise). Evidently, making $\alpha_2$ smaller would eventually deprioritize the goal of maintaining the ICU occupancy below the limit. Without loss of generality, we will use $\alpha_2=1e9$ in the remainder of this paper. In contrast, we found the reward function to be highly sensitive to variation in $\alpha_3$. For  $\alpha_3>10^{-4}$, the relative cost (negative reward) of imposing control becomes prohibitive and leads to one of the extreme intervention strategies: Suppression policy to end the endogenous transmission at the earliest possible time and avoid imposing lengthy control measures; or a no-intervention policy which plainly leads to the minimum relative control cost. In practice, the inclination for a specific intervention strategy depends on the policy maker's priorities.  We observed  pronounced variation in the optimal policies and resulting ICU occupancy trajectories for smaller values of $\alpha_3$ (compare the first and third columns, Fig. S3). In Figure S4, we  demonstrate this variation for each parameter set and across the values of  $\alpha_3$. As shown in Fig. S4A, values of $\alpha_3$ smaller than $10^{-4}$ result in greater \textit{Cumulative herd immunity reward}. Thus,  when the relative cost of control is  modest, the optimal policy function will tend to maximize the reward by increasing the number of individuals removed from the susceptible pool, which in turn leads to greater \textit{Cumulative control reward} (Fig. S4B) and longer epidemic duration (Fig. S4C). Therefore, among the tested values, $\alpha_3=1e-4$ represents the middle ground between no-intervention and suppression policies, and is the value that we have used in the rest of this paper.\\

\subsection{No-intervention policy, uniform intervention policy and optimal policy }

\begin{figure}[!hbt]
\centering
\includegraphics[width= 0.8 \linewidth]{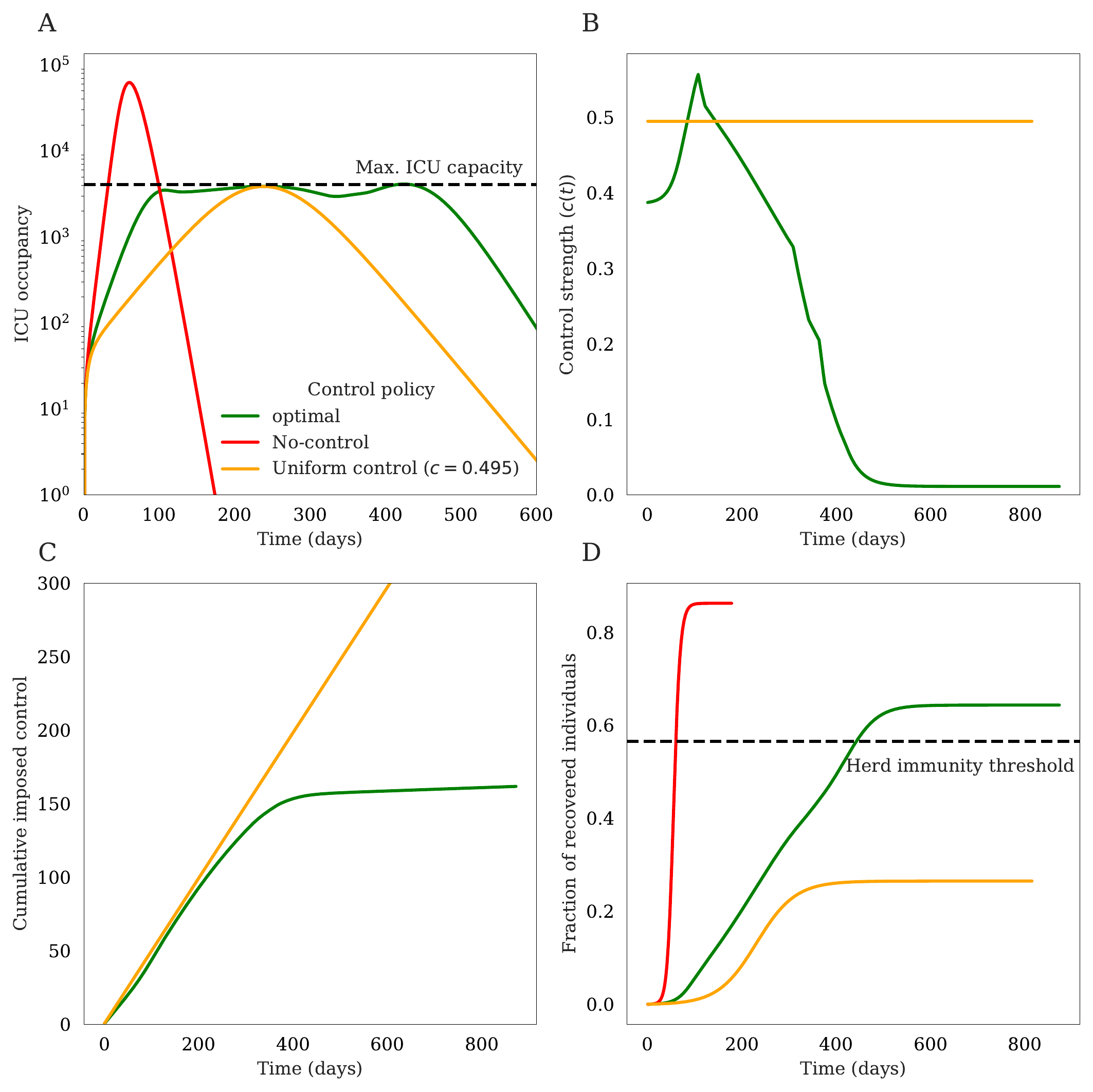}
\caption{{\bf No-intervention policy, Uniform intervention policy and optimal policy}
The figure presents the (A) ICU occupancy (B) Control strength (C) Cumulative imposed control and (D) recovered individuals for three different policies: No-intervention policy, Uniform intervention policy and optimal policy.}
\label{no_control_uniform_vs_optimal}
\end{figure}

 Figure \ref{no_control_uniform_vs_optimal} presents a comparison between the optimal intervention policy identified via our neuroevolution algorithm, a uniform intervention policy and no-intervention policy. The uniform intervention policy is implemented by imposing a constant reduction in transmission throughout the epidemic, $c(t)=c_u$. The value of control strength, $c_u$, is estimated such that the peak ICU occupancy tangents the maximum capacity. Figure ~\ref{no_control_uniform_vs_optimal}A depicts the ICU occupancy trajectories of these three policies. As expected, the no-intervention policy leads to ICU burdens well beyond the threshold capacity for more than two months (67 days). The other notable observation is the difference between the optimal and uniform policies in managing the ICU burden: the optimal policy maintains the ICU occupancy near the maximum capacity throughout the epidemic, but not  beyond it. Figure ~\ref{no_control_uniform_vs_optimal}B depicts the implemented control strength in time for optimal and uniform policies. Except for a period of time less than 10 weeks at the onset of the epidemic, the control strength of the optimal policy is below the uniform intervention policy. The difference in the imposed control between two policies is better illustrated by Figure ~\ref{no_control_uniform_vs_optimal}C, where a widening gap between the cumulative imposed control of the two policies emerges after day 200. In Figure ~\ref{no_control_uniform_vs_optimal}D, we present the recovered individuals for each policy. Unlike the optimal policy, the final fraction of recovered individuals in the uniform intervention policy case is well below the theoretical herd immunity threshold. This suggests that the any reduction in the control strength, could lead to another epidemic wave given the large fraction of susceptible individuals.

\subsection{The sooner the better}

\begin{figure}[!hbt]
\centering
\includegraphics[width= 0.65 \linewidth]{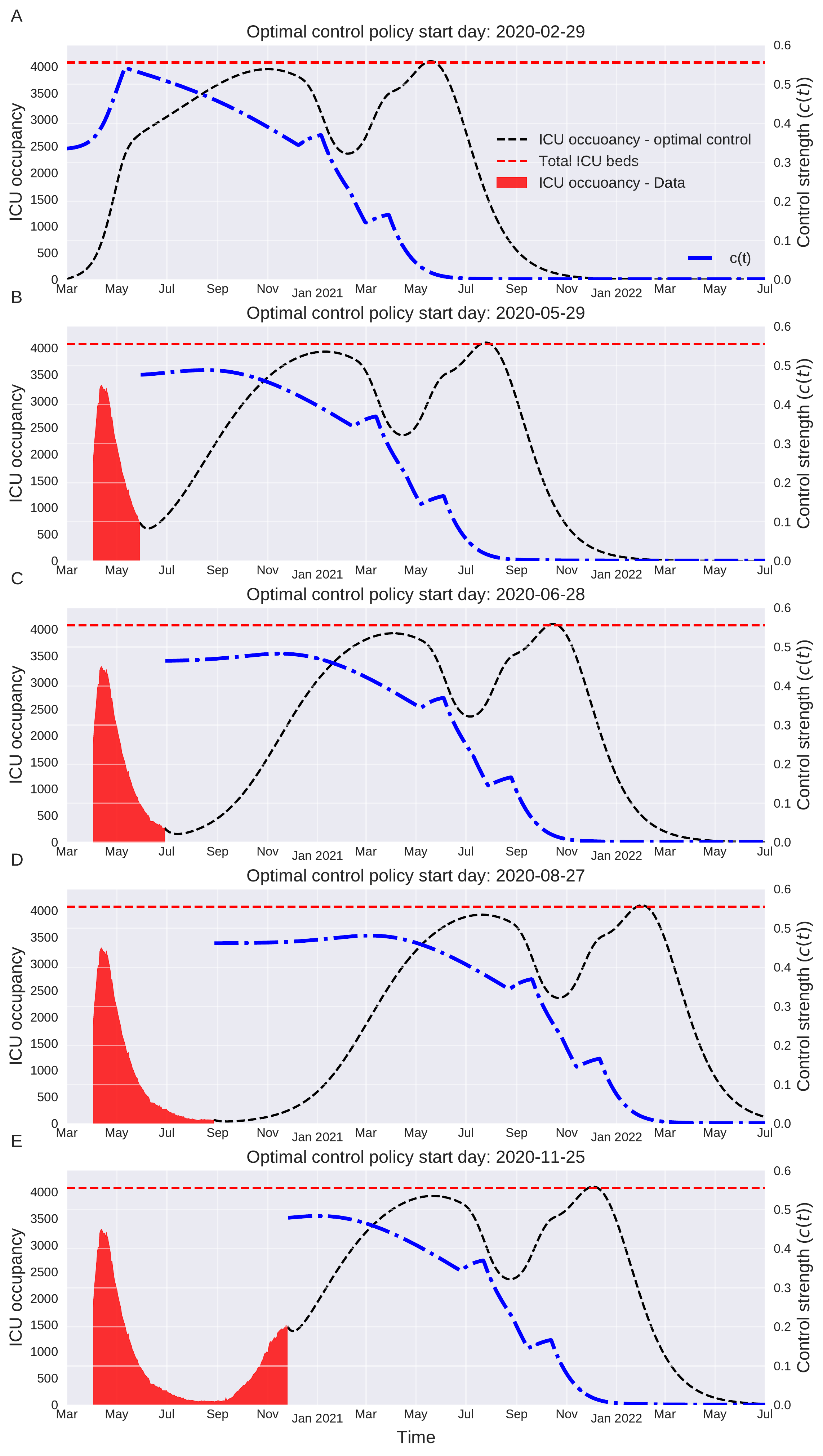}
\caption{{\bf Optimal intervention policy at different stages of epidemic}
The figure depicts the optimal intervention policy starting at different stages of epidemic. For each scenario, the number of susceptible, exposed, infectious and recovered individuals is estimated from a $SEIRH$ model fitted to the UK fatality data and used as initial condition to derive the optimal intervention policy.}
\label{control_scenarios}
\end{figure}

We have estimated the optimal intervention policy initiated at different stages of the epidemic, as shown in Figure \ref{control_scenarios}. Each scenario corresponds to a particular start date for the roll out of the optimal intervention policy. Figure \ref{control_scenarios}A depicts the scenario in which optimal intervention policy starts on March 1st, which coincides with a surge in cases in the UK. The optimal intervention policy starts with $c(t) = 0.33$ (a 33\% reduction in transmission rates) and is gradually increased to $c(t) = 0.54$ by mid-May.  The control strength tapers off to 0 by June 2021. This scenario leads to two peaks in ICU occupancy, in November 2020 and June 2021. Figures~\ref{control_scenarios}B-E depict the optimal intervention policy starting at intermediate stages of the epidemic. As mentioned above, we estimated the initial conditions for each scenario by fitting our $SIER$ model to fatality data using particle filtering, a Monte Carlo likelihood estimation algorithm for hidden state-space dynamical systems~\cite{doucet2009tutorial}. Comparing the optimal intervention policy curves in different scenarios depicts how implementing transmission reduction measures at earlier stages of the epidemic will eventually shorten the epidemic: The termination of optimal intervention policy is delayed from June 2021 (in Figure \ref{control_scenarios}A) to February 2022 (in Figure \ref{control_scenarios}D). The only exception is Figure \ref{control_scenarios}E, in which the optimal intervention policy terminates slightly sooner than in Figure \ref{control_scenarios}D. This is most likely due to the emergence of new variants with higher transmissibility~\cite{Davies2021EstimatedEngland.} which gave rise to a faster depletion of the susceptible pool than accounted for in our model.\\

\begin{figure}[!hbt]
\centering
\includegraphics[width= 0.8 \linewidth]{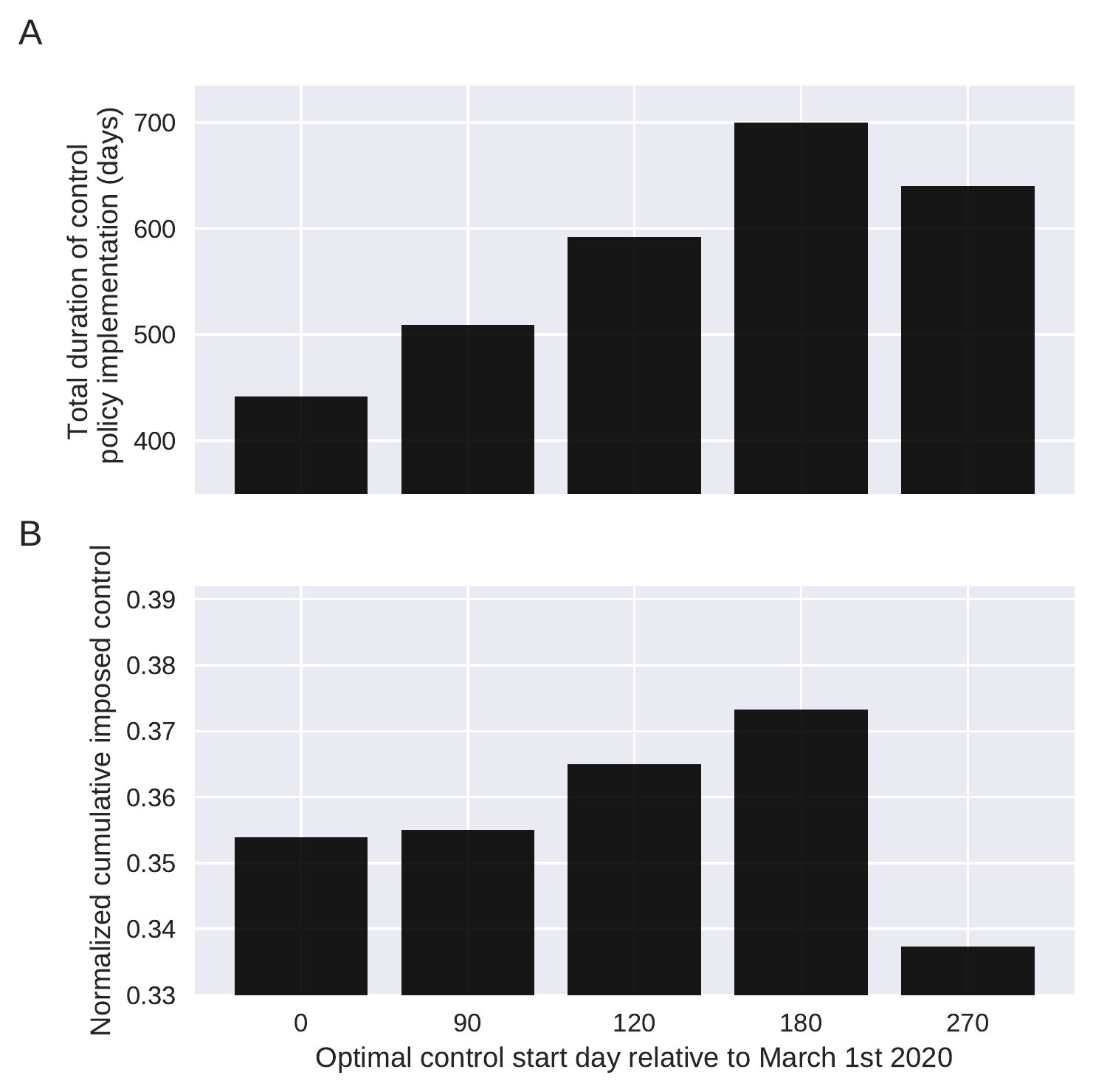}
\caption{{\bf Implementing the optimal intervention policy will reduce the overall impact of control measures}
the \textit{Total duration of intervention policy implementation} and \textit{Cumulative imposed control} for different scenarios. The \textit{Total duration of intervention policy implementation} represents the time period between March 1st 2020 and termination date of intervention policy for each scenario. The \textit{Cumulative imposed control} is obtained by adding up the implemented control strength ($c(t)$) in each day, divided by total number of days with $c(t)>0$ for each scenario.}
\label{cumulative_control}
\end{figure}

To better illustrate the importance of implementing early control measures, we have demonstrated the \textit{Total duration of intervention policy implementation} and \textit{Cumulative imposed control} for different scenarios in Figure \ref{cumulative_control}. The \textit{Total duration of intervention policy implementation} represents the time period between March 1st 2020 and the termination date of intervention policy for each scenario. The \textit{Cumulative imposed control} is obtained by summing the daily implemented control strength ($c(t)$), divided by total number of days with $c(t)>0$ for each scenario. As shown in Figure \ref{cumulative_control}A, the \textit{Total duration of intervention policy implementation} increases from 442 days in the first columns to 700 days in the last one. Figure \ref{cumulative_control}B also confirms the fact that implementing the optimal intervention policy from earlier stages of epidemic would reduce the overall required control measures. Note that depicted \textit{Cumulative imposed control} values do not include the actual imposed control strength ($c(t)$) before the start of optimal intervention policy and adding those values would only widen their differences. Also, the \textit{Cumulative imposed control} is a linear measure of overall imposed control, however, the actual economic cost would not necessarily change linearly with duration and strength of imposed intervention policy.\\

\subsection{Finding the balance}

\begin{figure}[!hbt]
\centering
\includegraphics[width= 0.8 \linewidth]{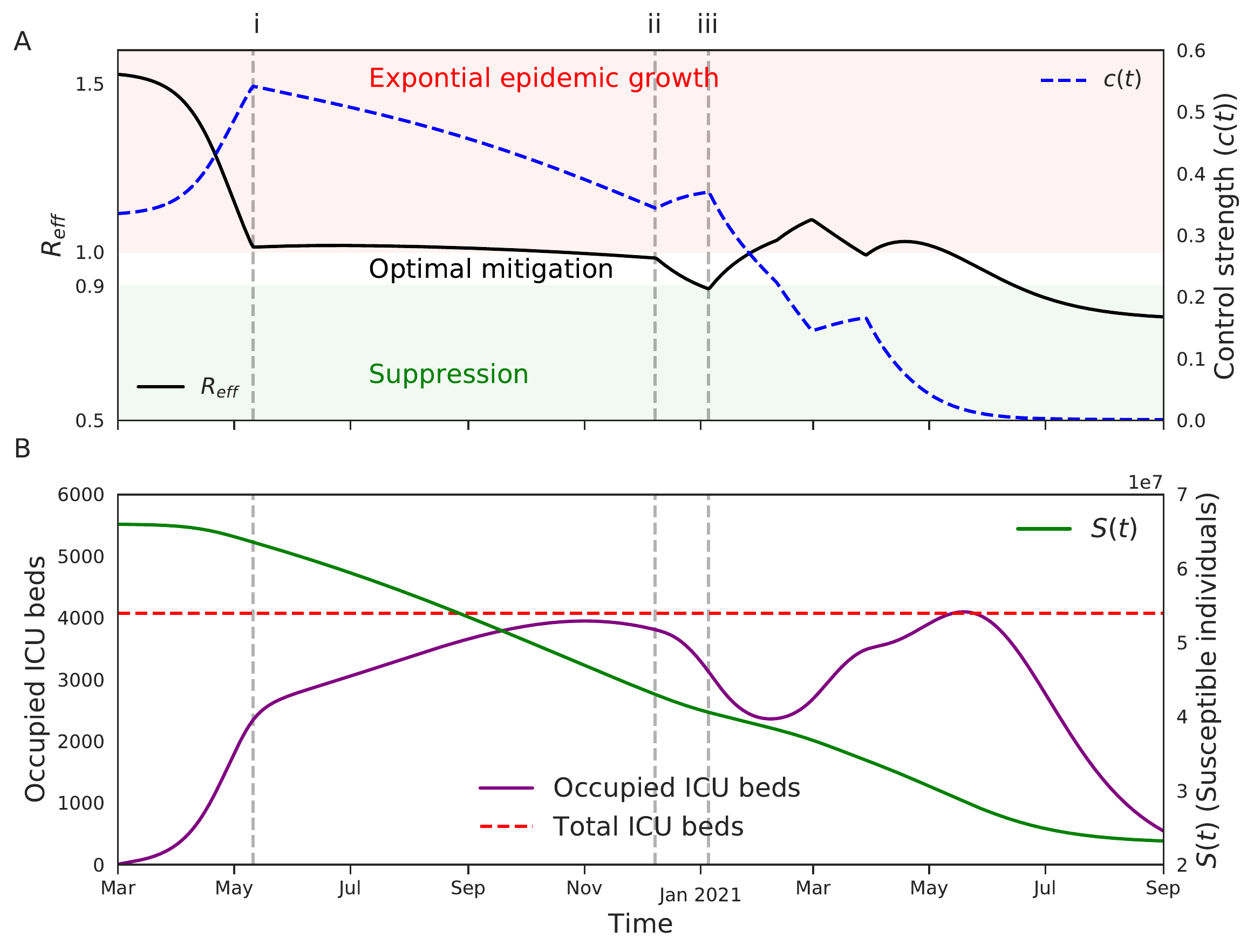}
\caption{{\bf The optimal intervention policy maintains the effective reproductive ratio ($R_{eff}$) close to 1:}
The figure displays the changes in effective reproductive ratio when implementing the optimal intervention policy. The control strength ($c(t)$) is sharply increased at early stages of epidemic to stall the epidemic growth and keep healthcare capacity from being overwhelmed. The $R_{eff}$ is maintained close to 1 by gradually reducing the $c(t)$ as the size of susceptible pool shrinks. Once the value of $R_{eff}$ reaches below 0.9, $c(t)$ is increased to sustain the transmission in the population, while keeping the occupied ICU beds below the maximum capacity.}
\label{Reffective_vs_policy}
\end{figure}

Figure \ref{Reffective_vs_policy} paints an overall picture of how the optimal policy fine tunes the transmission rates to sustain  endogenous transmission in the population without overburdening the ICU capacity.  Figure \ref{Reffective_vs_policy}A demonstrates the variation of effective reproductive ratio ($R_{eff}$) throughout the epidemic (black line), the control strength is also shown (blue dashed line).  At the onset of the epidemic,  $R_{eff}$ is instantly reduced to 1.52 from 2.3 by imposing a 0.33 reduction in contact rates ($c(t)=0.33$) and further decreased to $R_{eff} \approx 1 $  by mid-may (point i) to stall the epidemic growth. From point i to point ii, The $R_{eff}$ is maintained close to 1 to maintain the ICU occupancy close to the maximum capacity. At this point, $c(t)$ is slightly increased which leads to a sharp decrease of $R_{eff}$ to 0.89 in point iii. This is followed by a steep decrease in $c(t)$ to bring the $R_{eff}$ above 1 to sustain the transmission. To summarize, the optimal mitigation policy is achieved by finding the balance between two extreme scenarios: Suppression policy which aims to stall the endogenous transmission in the population, and "No-intervention" which leads to exponential epidemic growth and the overburdening of healthcare capacity.

\section{Discussion}

More than eighteen months into the SARS-CoV-2 pandemic, it is becoming increasingly clear that countries that implemented suppression strategies early on experienced greater success in managing both the public health and economic burden of the epidemic \cite{hassan2021hindsight,dong2020interactive,kochanczyk2021pareto}. However, such strategies work best when employed early in the epidemic, when number of cases is relatively small. Moreover, in countries where government-imposed restrictions are not well received by the public, implementation of such policies will be challenging. Looking back at the early stages of the epidemic, our work provides a dynamic mitigation strategy that sustains the community transmission without overwhelming the healthcare capacity.\\

A number of previous studies on optimal non-pharmaceutical interventions have used quadratic cost expressions for the control term in the cost function \cite{lee2010optimal,richard2021age,djidjou2020optimal}. This is mainly because when the cost function is quadratic with respect to the control, the differential equations arising from the necessary conditions for an optimal control have a known solution. Other functional forms frequently provide difficult-to-solve systems of differential equations. To circumvent this, we employed a neuroevolution algorithm which enabled us also to explore non-quadratic functions. The neuroevolution algorithm was used to train a policy function that takes the epidemiological state of population (the numbers of susceptible, exposed, infectious and recovered individuals) on each time day and provides the corresponding control strength. We defined a multi-objective reward function to account for three conflicting goals: Sustain the transmission to achieve herd immunity when suppression is not feasible, maintaining the ICU occupancy below the maximum capacity and imposing minimum possible control measures to reduce the contact rates. A relative weighting parameter was assigned to corresponding terms of each of these objectives in the reward function. The sensitivity analysis indicated that the resulting policy function is highly sensitive relative weighting of the control term and found a optimal range of of values for it.  We chose United Kingdom as our target population and fitted an $SEIRH$ model to fatality data to estimate the initial conditions at different stages of the epidemic. 
\\

The optimal intervention policy confirmed the importance of early interventions to reduce the contact rates in the population, as highlighted in the previous studies \cite{djidjou2020optimal,rowthorn2020cost}. An initial 34\% reduction in transmission at the onset of the epidemic, gradually increasing to ~50\% in the next 10 weeks is required to bring the $R_{eff} $ near 1. After that, the restrictions are constantly decreased as the the size of susceptible pool diminishes. The association between the control strength and the size of the susceptible pool (except the first initial 10 weeks) highlights the importance of reliable and widespread serosurveys in order to inform policy decision making.\\

A key component of our neuroevolution algorithm is the assumption that the full epidemiological state of the population is observable at each time step. In reality, however, the observable data provide an incomplete and potentially biased picture of epidemiology since they are based on reported incidence, hospitalization and fatality data in addition to seroprevalence surveys. Besides  assuming complete epidemiological information, our approach also assumed that the optimal intervention policy is implemented in deterministically; that is, the output action is perfectly implemented at each time instant and the resulting new state given the corresponding action is always the same - something that is not practical \cite{saeidpour2019probabilistic,saeidpour2018parameterized}. An important next step in this area would be to extend our novel framework to identify the optimal intervention strategies with hidden states in a stochastic setting. Furthermore, while this study addresses the optimal reduction in the contact rates over time, the economic cost and effectiveness of various non-pharmaceutical intervention mechanisms \cite{liu2021impact,courtemanche2020strong} to achieve the optimal policy reduction requirements must also be examined.

\section{Acknowledgments}
Research reported in this publication was supported by the National Institute Of General Medical Sciences of the National Institutes of Health under Award Number R01GM123007. The content is solely the responsibility of the authors and does not necessarily represent the official views of the National Institutes of Health.

\bibliographystyle{unsrt}  
\bibliography{templateArxiv}

\clearpage

\setcounter{page}{1}
\section*{Supplementary information}

\subsection*{Comparison of PMP and Neuroevolution optimal policies}
We have derived the optimal control solution via Pontryagin’s maximum principle (PMP) and compared the results with neuroevolution optimal policy in Figure S1.

\renewcommand{\thefigure}{S\arabic{figure}}

\begin{figure}[!hbt]
\label{PMP_VS_Neuro}
\includegraphics[width=\linewidth]{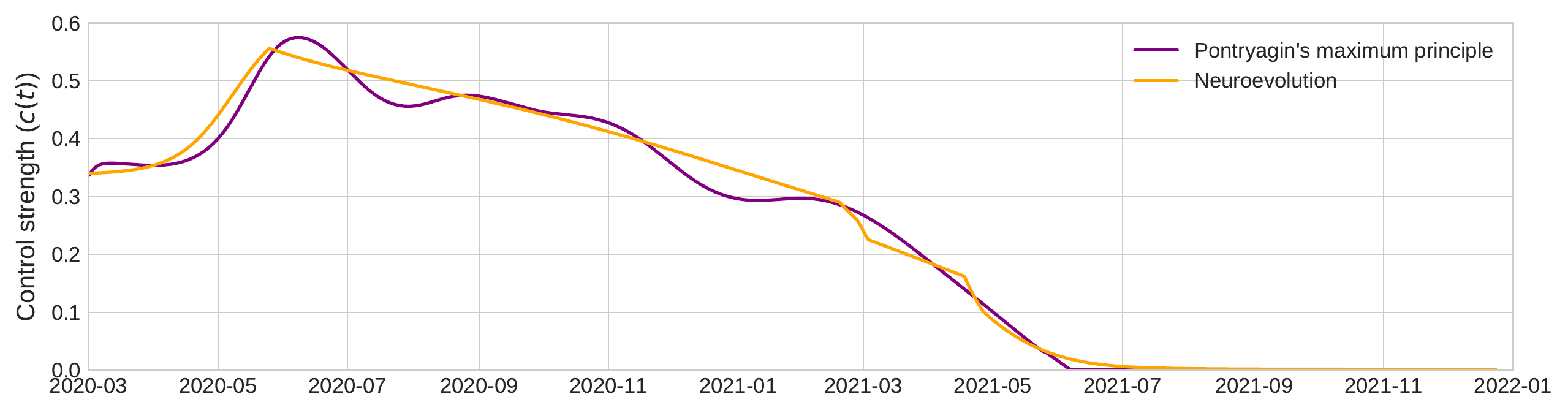}
\caption{{\bf PMP vs. Neuroevolution intervention policy}}

\end{figure}

\clearpage

\subsection*{SEIRH model fit to the fatality data}

Here we present the SEIRH model fitted on the daily fatality data via particle filtering. The model parameters are described in Table 1 in the main text and the model was fitted to estimate the control strength $c(t)$. We used the fitted model to estimate the initial conditions at different stages of the epidemic for optimal control analysis. 

\renewcommand{\thefigure}{S\arabic{figure}}
\begin{figure}[!hbt]
\includegraphics[width=\linewidth]{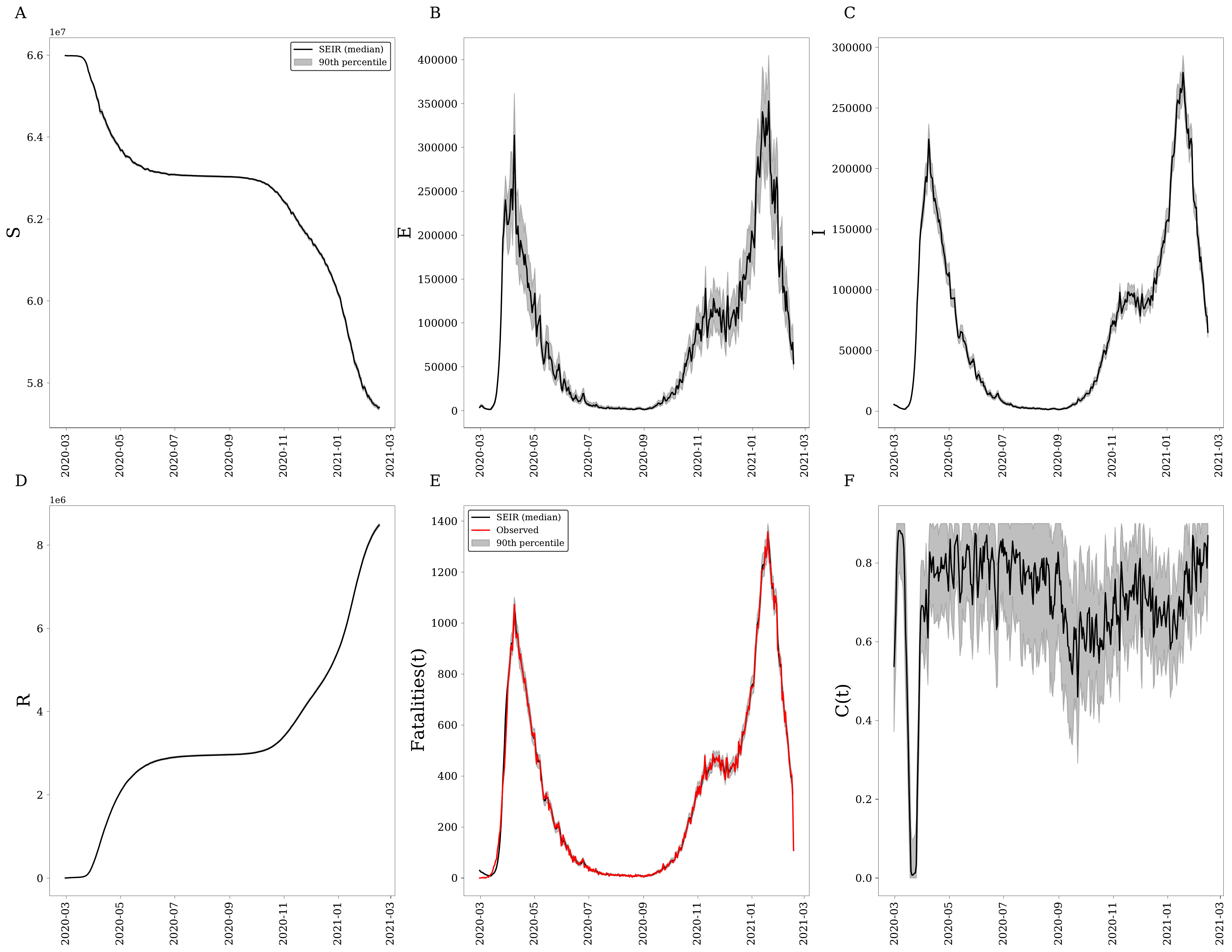}
\caption{{\bf SEIRH model fitted to the daily fatality data}
The figure shows the number of (A) susceptible (B) exposed (C) infectious (D) recovered classes from the fitted SEIRH model. The number of daily fatalities from the data and the model is shown in panel (E). Panel (F) depicts the estimated control strength ($c(t)$). In each panel the black line corresponds to the median of filtering distribution and the shaded area depicts the 90th percentile of filtered particles. The red line in panel (E) presents the fatality data.}
\label{SEIR_fit}
\end{figure}

\clearpage

\subsection*{Sensitivity analysis}
We carried out a sensitivity analysis to investigate the impact of relative weighting of each term in the reward function on the observed optimal policy outcome. This section presents the corresponding results.




\renewcommand{\thefigure}{S\arabic{figure}}
\begin{figure}[!hbt]
\includegraphics[width=\linewidth]{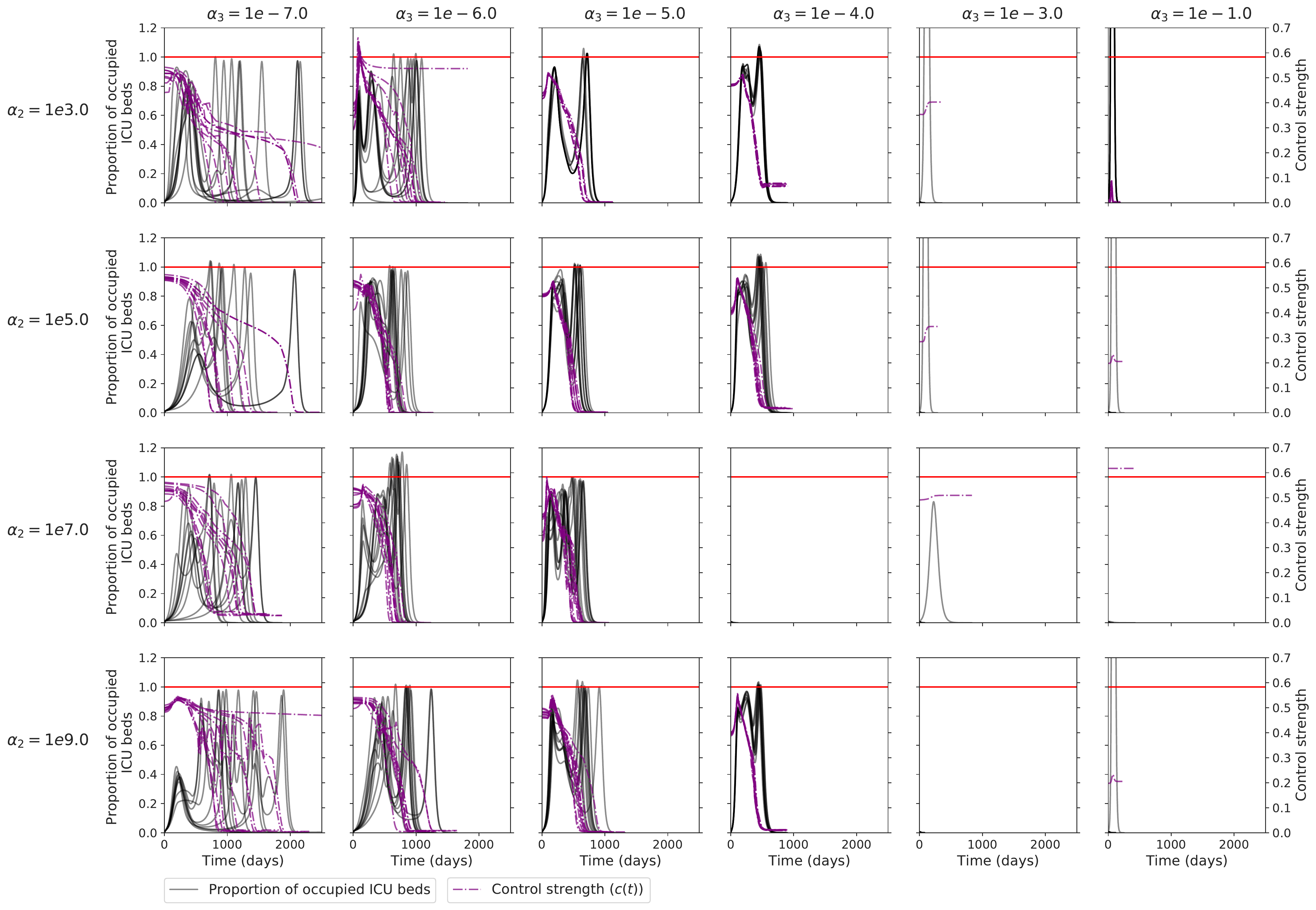}
\caption{{\bf Sensitivity analysis of reward function parameters}
The figure depicts optimal control policy and ICU occupancy trajectory of the 5 most elite agents for each $\{\alpha_2,\alpha_3\}$ combination.}
\label{sensitivity_grid}
\end{figure}

\renewcommand{\thefigure}{S\arabic{figure}}
\begin{figure}[!hbt]
\includegraphics[width=\linewidth]{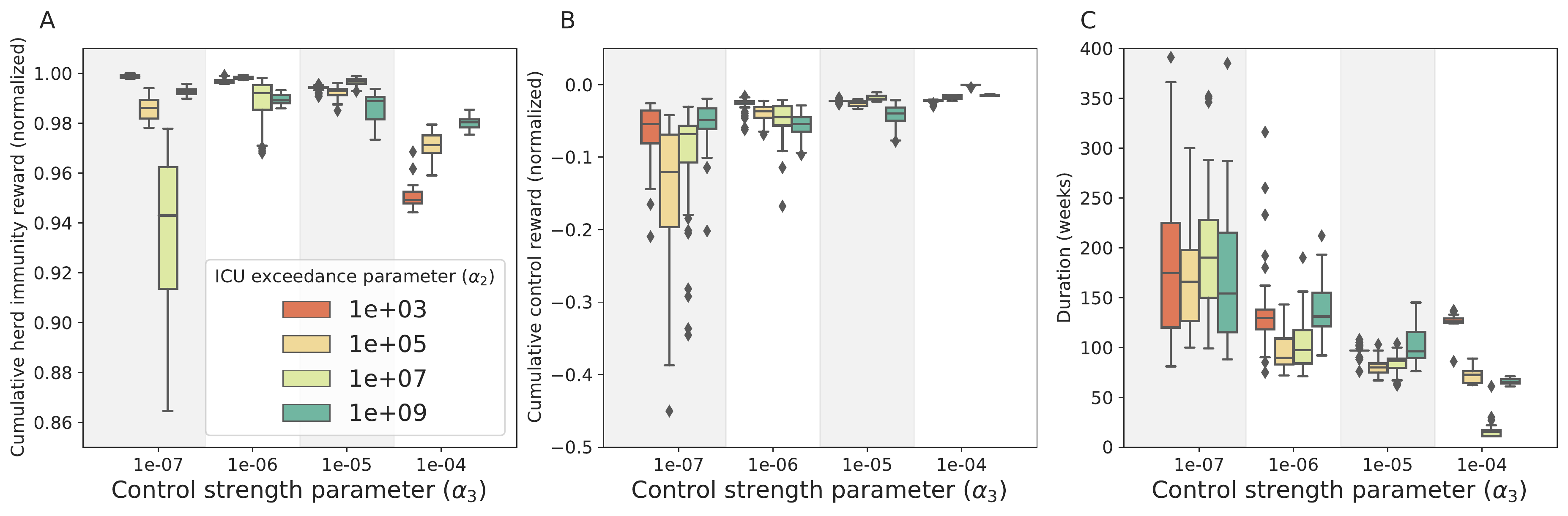}
\caption{{\bf The optimal control policy is mainly governed by weighting of control strength in the reward function.}
The top 50 policy functions for each $\{\alpha_2,\alpha_3\}$ combination is selected and used to reconstruct the epidemic trajectory. Panels denote the aggregated (A) Normalized cumulative herd immunity reward (B) Normalized cumulative control reward (C) Duration of imposing control measures for corresponding $\{\alpha_2,\alpha_3\}$ values.}
\label{sensitivity_summary}
\end{figure}

\clearpage

\end{document}